\documentclass[twocolumn,showpacs,preprintnumbers,amsmath,amssymb,prl,floatfix]{revtex4}
\usepackage{graphicx}
\usepackage{dcolumn}
\usepackage{bm}

\begin{document}

\title{Velocity distribution measurements in atomic beams generated using laser induced back-ablation}

\author{A. Denning, A. Booth, S. Lee, M. Ammonson, and S. D. Bergeson}
 \affiliation{Department of Physics and Astronomy, Brigham Young
 University, Provo, UT 84602, USA}

\date{\today}

\begin{abstract}
We present measurements of the velocity distribution of calcium atoms in an atomic beam generated using a dual-stage laser back-ablation apparatus.  Distributions are measured using a velocity selective Doppler time-of-flight technique.  They are Boltzmann-like with rms velocities corresponding to temperatures above the melting point for calcium.  Contrary to a recent report in the literature, this method does not generate a sub-thermal atomic beam.
\end{abstract}


\maketitle

\section{Introduction}

A recent publication reported a sub-thermal atomic beam generated by laser induced back-ablation \cite{alti05}.  In this method, a thin film that has been deposited onto a transparent substrate is ablated by laser illumination through the substrate.  Atomic beams generated this way can be collimated or focused depending on the geometry of the ablation laser \cite{kallen89, bullock99}.  This has obvious advantages for spectroscopy and atomic physics experiments.

The sub-thermal velocities reported in Ref. \cite{alti05} are surprising because there is no known mechanism to explain this.  Ejected plumes from laser ablation sources have been studied extensively \cite{tallents80, albritton86, dreyfus86, wang91, kools91, rupp95, chichkov96, zhigilei97, hansen97, toftmann00}.  While the details of ablation using ns-duration lasers are complicated, the general physical principles are well known.  Laser light is absorbed by a target.  The laser pulse energy raises the local temperature to the melting point and atoms evaporate from the surface.  The temperature of the ablated material is determined by the thermal conductivity of the target material and the laser pulse duration.  For very high intensity lasers, atoms in the ablated plume can be reach high ionization states.  The electron transport properties can be nonlocal.  In high density plasmas, significant recombination can also occur.  Depending on the density and experimental conditions, shock fronts can also form in the expanding plasma.

Apparently, there are no velocity distribution measurements for atomic beams generated via back-ablation.  Early work on back-ablated beams \cite{kallen89} concentrated on angular distributions and assumed the ion temperature was equal to the target melting temperature.  Somewhat more recent work interferometrically measured the plume edge velocity in high-intensity picosecond laser back-ablated plasmas \cite{bullock99}.  The recent work of Ref. \cite{alti05} used a time-of-flight laser deflection method to deduce the beam velocity.

In this paper we report new measurements of the velocity distribution in a calcium atomic beam generated by a dual-stage back-ablation atomic beam source. We measure the distribution using a velocity-selective Doppler time-of-flight method, similar to previously published work (see, for example, Ref. \cite{dreyfus86}).  We find that at moderate intensities of the back-ablation laser, the distribution is Boltzmann-like.  At the lowest intensities, when the calcium film is not completely ablated, the rms velocity distribution never falls below a thermal distribution at the melting temperature.

\section{Experiment}

\begin{figure}
\includegraphics[angle=270,width=3.2in]{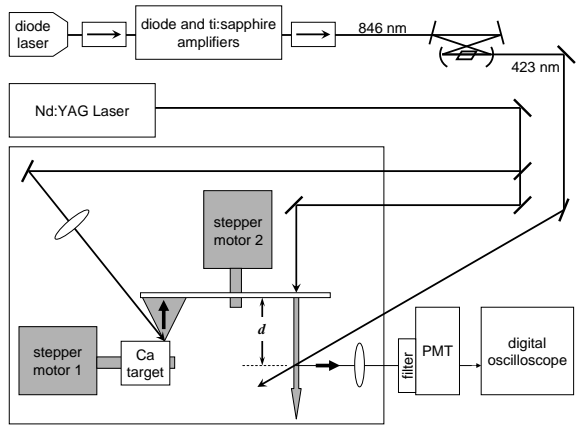}
\caption{\label{fig:schematic} A Schematic diagram of our dual-stage laser ablation experiment.  An Nd:YAG laser beam is divided into two beams.  One is used to ablate a thin calcium film onto a sapphire substrate.  The other is used to generate the collimated calcium beam using back-ablation.  A single-frequency probe laser at 423 nm excites specific velocity classes in the atomic beam, depending on the laser frequency. The ablation targets and associated optics are placed inside a vacuum system that is maintained using a turbo-molecular pump. }
\end{figure}

A schematic diagram of our experiment is shown in Fig. \ref{fig:schematic}.  The two-stage back-ablation source is similar to Ref. \cite{kallen89}.  The front surface of a transparent sapphire substrate is coated by a thin film of calcium using regular laser ablation.  This thin film is then ablated off of the substrate by illuminating the thin film through the back of the sapphire disk using another high intensity laser beam.

A 532 nm Nd:YAG laser with a 3 ns duration pulse is used for ablation.  The laser pulse is divided into two beams.  One is focused using a 90 mm focal length lens onto a calcium target.  The target is mounted on a stepper motor.  It is rotated and advanced in such a way that each laser shot ablates a new target region.  Target is advanced back and forth repeatedly and the experiment is run for several hours and the surface of the ablated target is not smooth.

The ablated metal deposits a thin film on a nearby rotating sapphire disk.  The thickness of the film can be adjusted by attenuating the ablation laser or by controlling the rotation speed of the sapphire disk.  The disk has a 10 cm diameter and is rotated at a frequency of $\sim 0.1$ Hz.  The calcium target is located 1.5 cm from the disk, and the ablated plume coats an area of approximately 3 cm diameter near the outer edge of the disk.  Contrary to Ref. \cite{kallen89}, the sapphire disk is continuously loaded.  The thickness of the calcium film is determined by the angular velocity of the disk and the number of laser shots that coat the disk with ablated atoms while the disk covers an arc equivalent in length to the width of the ablating beam.  The Nd:YAG laser operates at 10 Hz.  The calcium film thickness results from approximately 30 laser shots as the disk rotates through the ablated plume.  The second ablation laser beam is weakly focused to a Gaussian waist of 3 mm at the sapphire disk and back-illuminates the calcium thin film.  At full power (30 mJ), the calcium film is completely ablated.

A probe laser beam at 423 nm crosses the calcium beam at an angle of 53$^{\rm o}$ at a distance $d$ from the sapphire disk (see Fig. \ref{fig:schematic}).  The probe laser beam is generated by frequency doubling an injection-locked ti:sapphire laser, and is described elsewhere \cite{cummings03}.  The bandwidth of this laser is less than 1 MHz, and the laser frequency can be swept across the Doppler-broadened $4s^2 \; ^1S_0 \rightarrow 4s4p \; ^1P_1^{\rm o}$ calcium resonance line.  Not shown in this figure is a calcium vapor cell in which we measure the calcium resonance transition using saturated absorption spectroscopy.  This measurement provides an important check on the initial frequency of the probe laser beam.

The fluorescence signal is collected using a lens, measured using a 1P28 photomultiplier tube (PMT), and digitized and averaged using a fast oscilloscope.  Special care is taken to ensure the linearity of the PMT response and that the detection bandwidth is high enough not to compromise the fluorescence signal.  By imaging the intersection of the atomic and laser beams onto the PMT photocathode through an aperture, we optimize the fluorescence signal strength and minimize the scattered Nd:YAG laser light.

\begin{figure}
\includegraphics[width=3.2in]{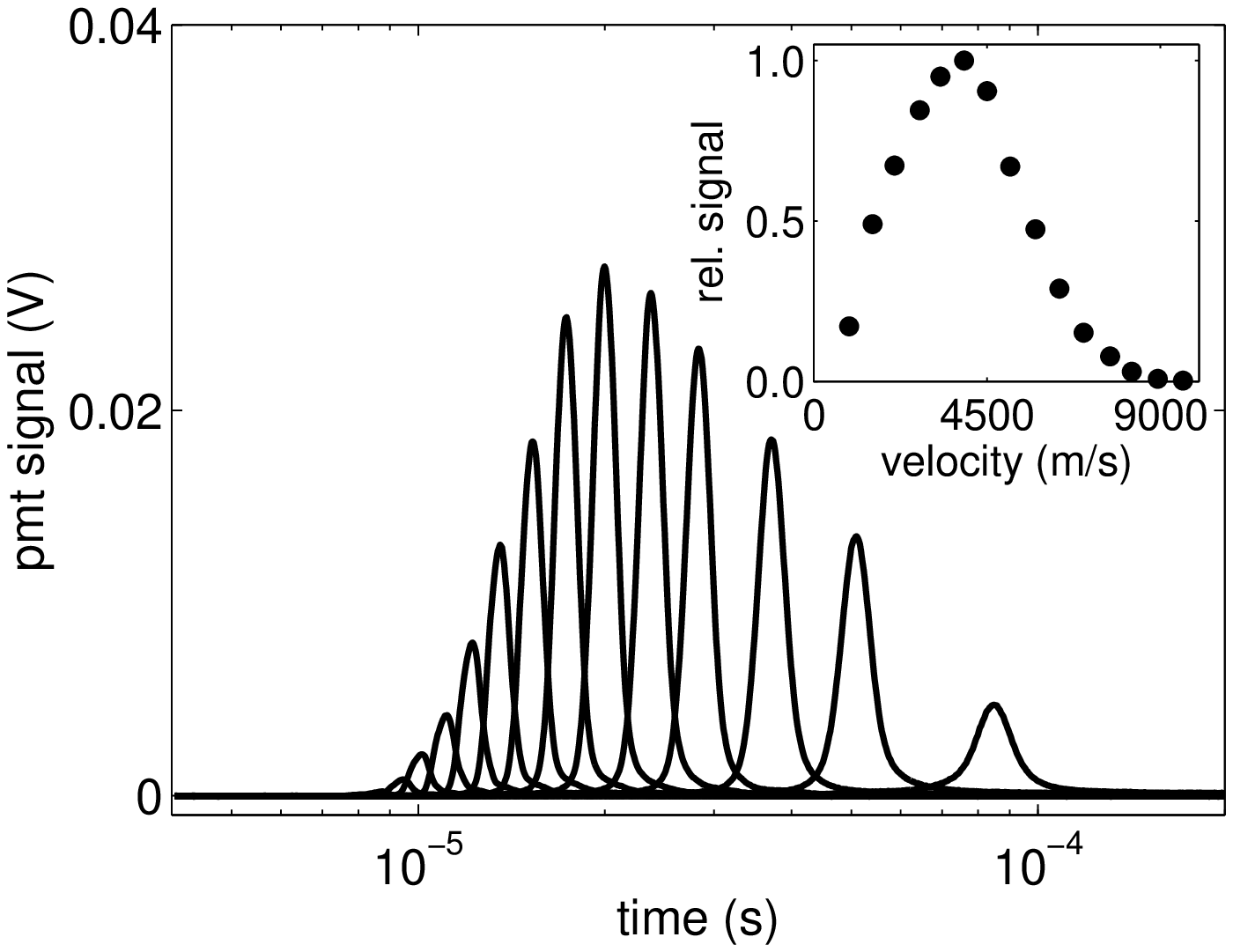}\\
\includegraphics[width=3.2in]{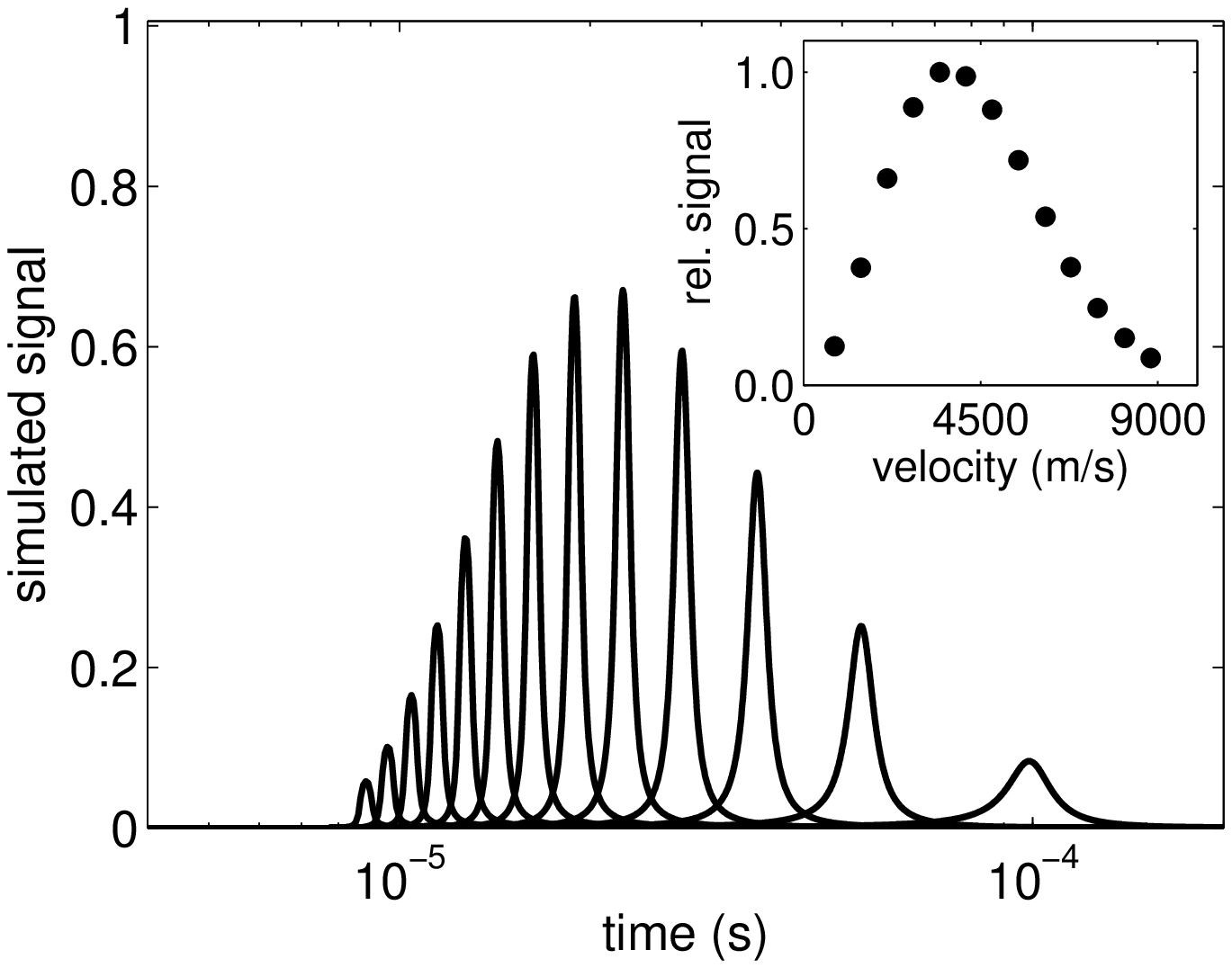}
\caption{\label{fig:signal} Measured (top panel) and simulated (bottom panel) laser-induced fluorescence signals from atoms in the atomic beam for a range of different probe laser beam detunings and $d=7.8$ cm.  As the probe laser beam frequency detuning increases, the laser interacts with faster atoms and the fluorescence signal peak occurs earlier in time. The inset to each plot is the velocity distribution in the atomic beam as determined by the fluorescence signal.}
\end{figure}

The wavelength of the probe laser beam is monitored using a wavemeter.  For each probe laser wavelength setting we measure the laser-induced fluorescence and average over 100 laser shots.  Typical fluorescence curves are shown in Fig. \ref{fig:signal}.  As we increase the detuning of the probe laser beam from resonance, the laser beam interacts with faster atoms and the peak of the fluorescence signal occurs earlier in time.

\section{Analysis}

An atom in the atomic beam will fluoresce when it is Doppler-shifted into resonance with the probe laser beam.  The probability of fluorescing depends on both the atomic line shape and the velocity distribution.  For an individual atom moving at some velocity $v$, the fluorescence signal will be a Doppler-shifted Lorentzian:

\begin{eqnarray}
{\cal L} & = & \frac{\gamma/2\pi}{[\nu_L - \nu_0(1 + v\;\cos\theta/c)]^2 + \gamma^2/4} \\
& = & \frac{\gamma/2\pi}{(\Delta - v\;\cos\theta/\lambda)^2 + \gamma^2/4}, \label{eqn:lorentz}
\end{eqnarray}

\noindent where $\gamma$ is the full-width at half-maximum (FWHM), $\nu_L$ is the laser frequency, $\nu_0$ is the resonance frequency in the rest frame of the atom, $v$ is the atom velocity, $c$ is the speed of light, $\Delta = \nu_L - \nu_0$ is the laser detuning, $\theta$ is the angle between the atomic beam direction and the probe laser wave beam direction, and $\lambda$ is the transition wavelength.
For the 423 nm resonance transition in calcium, the FWHM is 35 MHz.
Other factors, such as power broadening of the atomic transition and divergence of the atomic beam, contribute additional width to the atomic transition.
In our experiment, we set the laser detuning $\Delta$ and measure the fluorescence as a function of time.

We have performed a simulation of our experiment.  At a time $t=0$, atoms with a particular velocity $v$ are launched from the sapphire disk.  At a time $t_0 = d/v$ the atoms encounter the probe laser beam.  If the width of the probe laser beam is $w$, the atoms spend a time $\tau = w/v$ in the laser beam.  For simplicity, we assume that atoms scatter photons at a rate given by Eq. \ref{eqn:lorentz} only while they are in the beam.  This is a simplification to the Gaussian spatial profile of the probe laser beam.  However, the intensity of the probe laser is approximately 10$\times$ the saturation intensity, making this flat-top approximation somewhat more realistic.  The fluorescence signal from each velocity class is multiplied by the probability of finding that velocity in the Maxwell-Boltzmann distribution, $\exp(-v^2/2v_{th}^2)$.  We launch atoms with a range of velocities and sum their fluorescence signal to simulate the total signal at a particular laser detuning $\Delta$.

Typical fluorescence data for $d=7.8$ cm are shown in the top panel of Fig. \ref{fig:signal} for a range of laser detunings.  As the laser detuning increases, the probe laser beam interacts with faster atoms and the fluorescence signal peaks at earlier times. The inset to each plot shows the derived velocity distribution.  These distributions are determined from the peak of the fluorescence signal at each laser detuning.  The velocity can be determined either from the wavemeter reading or from the time of flight.  These methods agree to within the uncertainties of the wavemeter data.

The bottom panel of Fig. \ref{fig:signal} shows the result of our simulation for $d=7.8$ cm.  As mentioned previously, beam divergence and power broadening contribute to the width of the fluorescence signal at a given laser detuning.  We approximate those effects by increasing the Lorentzian linewidth from the natural linewidth of 35 MHz to a power-broadened linewidth of 150 MHz.  The probe laser beam is approximately 50 mW at 423 nm with a Gaussian waist around 1.5 mm.
Based on our measurements of the fluorescence signals for different values of $d$, the atomic beam divergence appears to be less than 15 mrad.  This is consistent with earlier studies of atomic beams generated by back-ablation \cite{kallen89,bullock99}.

The general agreement between the experiment and the simulation in Fig. \ref{fig:signal} indicates that the velocity distribution in the atomic beam is generally Maxwellian. The thermal velocity $v_{th}=\sqrt{k_BT/m}$ in the simulation is 3000 m/s, corresponding to a temperature of $T = 44,000$ K $= 3.8$ eV.  We have repeated these measurements for lower back-ablation laser intensity.  We find that the rms velocity decreases, but so does the ablation efficiency. For our lowest intensity, the rms velocity appears to be 1000 m/s, corresponding to a temperature of $4,000$ K, well above the melting temperature.  In this limit, the back-ablation does not completely ablate the calcium film, meaning that the thickness of the film increases over time.

These data show that for all back-ablation intensities, the longitudinal velocity in the beam corresponds to temperatures well above the melting temperature of the ablated metal.  While the details of the model and the distributions can be argued, we see no evidence for a sub-thermal atomic beam as reported recently \cite{alti05}.

\section{Discussion}

Velocity-selective Doppler time-of-flight methods provide direct information about the velocity distribution in atomic beams.  In particular, it can be used to determine both the speed and direction of atoms that cross the probe laser beam.  In a compact beam apparatus, such as we use, this can be important.

\begin{figure}
\includegraphics[width=3.2in]{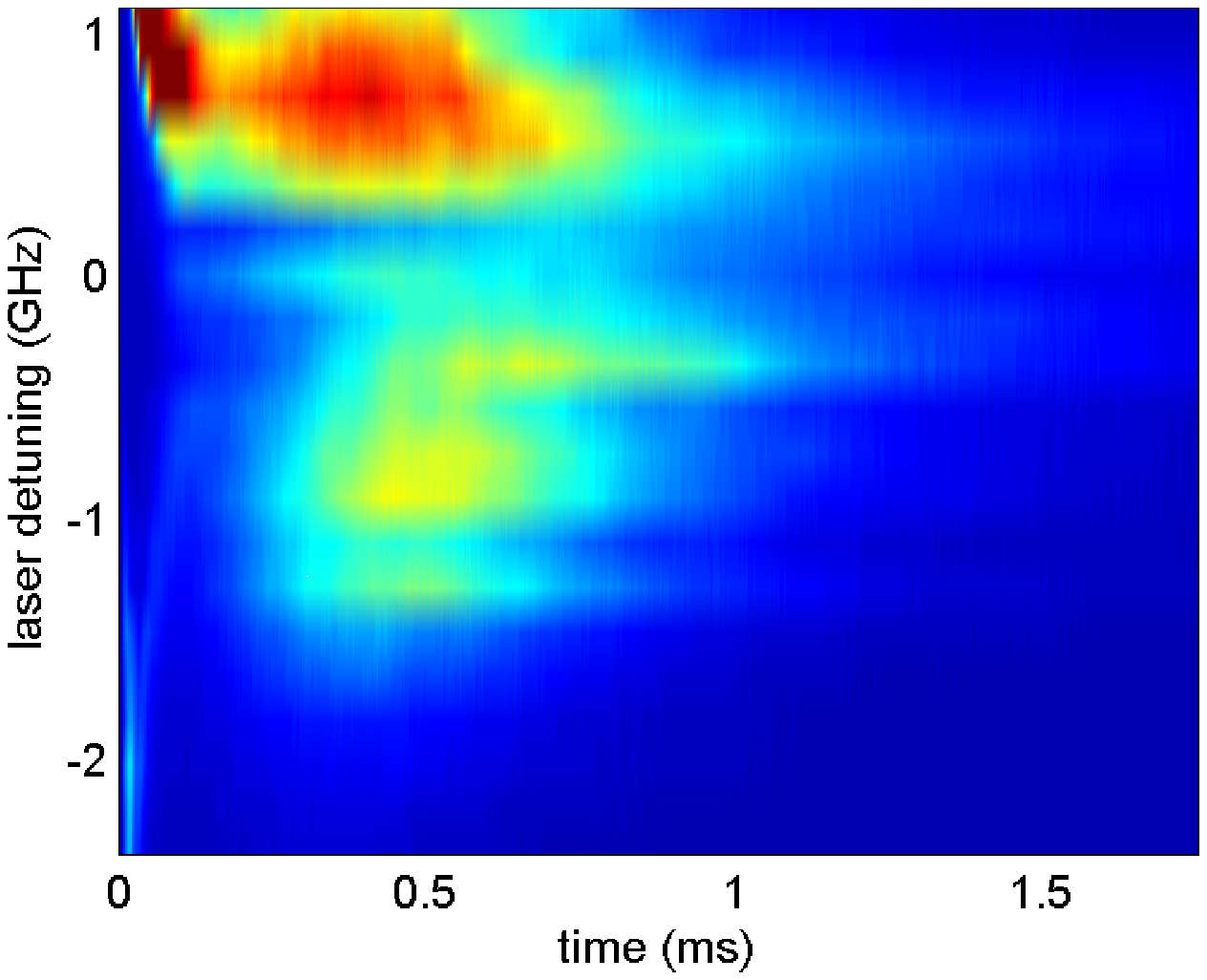}\\
\includegraphics[width=3.2in]{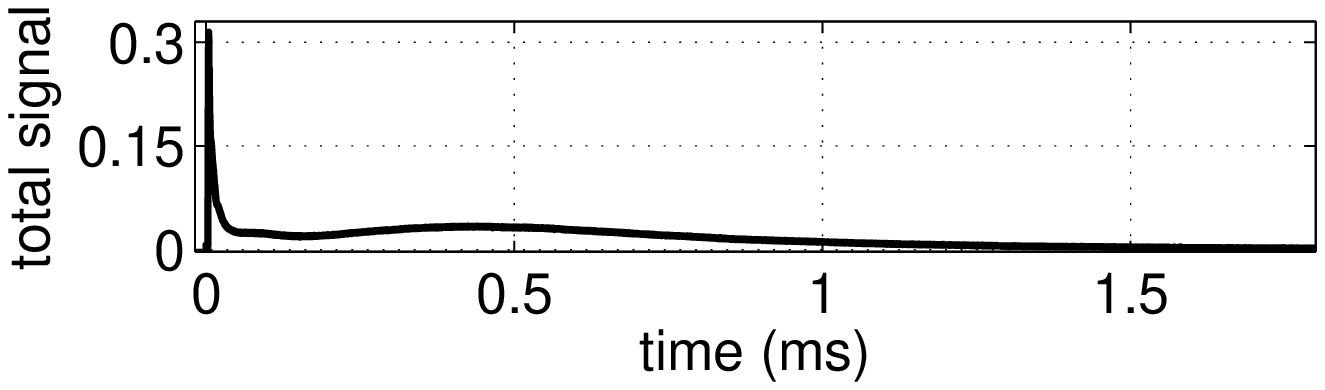}
\caption{\label{fig:bounce} (color) Measurements of the velocity distribution in an un-optimized experimental setup.  Top: False color plot of the fluorescence signal versus time and probe laser detuning.  Anomalously slow signals appear at small positive and negative probe laser beam detunings.  Anomalous signal also appears at large negative detunings (lower left portion of the image).  These signals are due to a poor experimental design (see text) and are dramatically reduced in a more carefully configured experiment.  Bottom: The total fluorescence signal summed over all laser detunings, from -2.5 GHz to +10 GHz.}
\end{figure}

In an early iteration of our experiment, we found features of the velocity distribution that did not agree with a Maxwellian distribution.  We measured anomalous fluorescence signals at negative and also small positive values of the laser detuning $\Delta$.  A representation of that data is shown in Fig. \ref{fig:bounce}.  As we studied the data and apparatus more carefully, we found evidence of interaction between atoms from the solid calcium target ablation and the thin film back-ablation.  It is possible that atoms moving in the backwards direction are generated by the transmitted back-ablation beam interacting with calcium atoms deposited on the beam block.  This could explain the weak and very fast signal component at very early times at large negative detunings.  The apparently slow atoms seen at small positive detunings could come from the interaction of the forwards and backwards traveling atomic beams, or from the interaction between atoms in the two different ablation sources.  We find that in a more carefully designed experiment, such as shown in Fig. \ref{fig:schematic}, in which the atomic beam path is clear from obstruction, the anomalous signal is significantly reduced.

Without a velocity-selective probe, it is impossible to accurately characterize the velocity profile of the beam. If we mistakenly assume that the late peak at 0.5 ms in the bottom panel of Fig. \ref{fig:bounce} corresponds to atoms generated by the back ablation laser at the disk, we derive an atomic velocity near 80 m/s and a temperature of 30 K.  However, more careful measurements show that such a conclusion is in error.

\section{Conclusion}

We report measurements of longitudinal velocity distributions of calcium atoms in an atomic beam generated using a two-stage back-ablation experiment.  We use a Doppler-selective time-of-flight method to determine the distributions.  Fluorescence signals are reproduced in a simple simulation assuming a Maxwellian distribution at a temperature well above the melting temperature.  We find no evidence for sub-thermal velocities as recently reported in the literature.  Moreover, we point out that there is no know physical mechanism for producing a sub-thermal atomic beam by laser ablation.

This work was supported in part by the Research Corporation, Brigham Young University, and by the National Science Foundation (PHY-0601699).


\begin{thebibliography}{99}

\bibitem{alti05}
K. Alti and A. Khare, Rev. Sci. Instrum. 76, 113302 (2005)

\bibitem{kallen89}
M. A. Kadar-Kallen and K. D. Bonin, App. Phys. Lett. 54, 2296 (1989)

\bibitem{bullock99}
A. B. Bullock and P. R. Bolton, J. Appl. Phys. 85, 460 (1999)

\bibitem{tallents80}
G. J. Tallents, Plas. Phys. 22, 709 (1980)

\bibitem{albritton86}
J. R. Albritton, E. A. Williams, I. B. Bernstein, and K. P. Swartz, Phys. Rev. Lett. 57, 1887 (1986)

\bibitem{dreyfus86}
R. W. Dreyfus, R. Kelly, and R. E. Walkup, Appl. Phys. Lett. 49, 1478 (1986)

\bibitem{wang91}
H. Wang, A. P. Salzberg, and B. R. Weiner, Appl. Phys. Lett. 59, 935 (1991)

\bibitem{kools91}
J. C. S. Kools, S. H. Brongersma, E. van de Riet, and J. Dieleman, Appl. Phys. B 53, 125 (1991)

\bibitem{rupp95}
A. Rupp and K. Rohr, J. Phys. D: Appl. Phys. 28, 468 (1995)

\bibitem{chichkov96}
B. N. Chichkov, C. Momma, S. Nolte, F. von Alfensleben, and A. T\"{u}nnermann, Appl. Phys. A 63, 109 (1996)

\bibitem{zhigilei97}
L. V. Zhigilei and B. J. Garrison, Appl. Phys. Lett. 71, 551 (1997)

\bibitem{hansen97}
T. N. Hansen, J. Schou, and J. G. Lunney, Europhys. Lett. 40, 441 (1997)

\bibitem{toftmann00}
B. N. Toftmann, J. Schou, T. N. Hansen, and J. G. Lunney, Phys. Rev. Lett. 84, 3998 (2000)

\bibitem{cummings03}
E. Cummings, M. Hicken, and S. Bergeson, Appl. Opt.  41, 7583 (2002)

\end{thebibliography}
\end{document}